\documentclass[12pt]{iopart}
\usepackage{iopams}
\usepackage{graphicx}

\begin{document}

\title[$^{57}$Fe M\"ossbauer effect in {\it R}Fe$_2$Zn$_{20}$]
{Study of $^{57}$Fe M\"ossbauer effect in {\it R}Fe$_2$Zn$_{20}$ ({\it R} = Lu, Yb, Gd).}

\author{Sergey L Bud'ko$^1$, Tai Kong$^1$, Xiaoming Ma$^{1,2}$, and Paul C Canfield$^1$}

\address{$^1$Ames Laboratory US DOE and Department of Physics and Astronomy,
Iowa State University, Ames, IA 50011, USA}

\address{$^2$Institute of Applied Magnetics, Key Laboratory for Magnetism and Magnetic Materials of the Ministry of Education, Lanzhou University, Lanzhou 730000, Gansu Province, China}

\begin{abstract}

We report measurements of $^{57}$Fe M\"ossbauer spectra for {\it R}Fe$_2$Zn$_{20}$ ({\it R} = Lu, Yb, Gd) from $\sim 4.5$ K to room temperature. The obtained  isomer shift values are very similar for all three compounds, their temperature dependence was analyzed within the Debye model and resulted in an estimate of the Debye temperatures of 450-500 K. The values of quadrupole splitting at room temperature change with the cubic lattice constant $a$ in a linear fashion. For GdFe$_2$Zn$_{20}$, ferromagnetic order is seen as an appearance of a sextet in the spectra. The $^{57}$Fe site hyperfine field for $T \to 0$ was evaluated to be $\sim 2.4$ T.

\end{abstract}


\submitto{\JPCM}

\maketitle

\section{Introduction}

The  {\it RT}$_2$Zn$_{20}$ ({\it R} = rare earth, {\it T} = transition metal) large family of dilute, rare earth containing intermetallic compounds was discovered almost two decades ago, \cite{nas97a} but only a decade later \cite{jia07a,tor07a} the physical properties of the members of this family started to be of interest for the broader physics community.  The studies of this family brought to light six new Yb-based heavy fermion compounds \cite{tor07a}, nearly ferromagnetic Fermi liquid behavior in YFe$_2$Zn$_{20}$  and LuFe$_2$Zn$_{20}$ \cite{jia07a,jia09a} and physics of dilute, local magnetic moments in nearly ferromagnetic Fermi liquid. \cite{jia07a,jia07b} 
For example the difference in the ordering temperatures and the nature of the magnetic ordered state in the Gd{\it T}$_2$Zn$_{20}$ ({\it T} = Fe, Ru, Os, Co, Rh, and Ir) was understood in the model of Heisenberg moments embedded in a nearly ferromagnetic, highly polarizable Fermi liquid, combined with the electronic band filling that strongly affect this polarizability. \cite{jia08a} Similar approach, with an addition of the effects of the crystal magnetic field, can be used to understand enhanced Curie temperatures in the {\it R}Fe$_2$Zn$_{20}$  ({\it R} = rare earth) series. \cite{jia09a,tia10a,isi13a,yaz14a}

The  {\it RT}$_2$Zn$_{20}$ family was mostly studied using bulk measurements of physical properties. Only few publications \cite{tia10a,wan10a,tam13a} reported results of the measurements by local probe techniques. One of such techniques is M\"ossbauer spectroscopy that allows for the determination of local electronic and magnetic properties. \cite{gut11a} Of the  {\it RT}$_2$Zn$_{20}$ family, $^{57}$Fe M\"ossbauer spectroscopy was recently used to study DyFe$_2$Zn$_{20}$ and YFe$_2$Zn$_{20}$. \cite{tam13a} The data presented for YFe$_2$Zn$_{20}$ at 299 K and 78 K indicated a paramagnetic state of this compound, for  DyFe$_2$Zn$_{20}$ the data at 8 K demonstrated presence of the hyperfine field consistent with a long range magnetic order, \cite{jia07a,jia09a,jia07b,jia08a}  however a complex combination one one unknown subspectrum and two main subspectra were used to fit the data. \cite{tam13a}

In this work we  use  $^{57}$Fe M\"ossbauer spectroscopy to study the properties of GdFe$_2$Zn$_{20}$, the compound that has the highest Curie temperature in the  {\it RT}$_2$Zn$_{20}$ family \cite{jia07a,jia09a,jia08a} and has no possible complications of crystal electric field effects since  Gd$^{3+}$ has Hund's rule ground state multiplet $J$ with zero angular momentum. Additionally, we perform measurements on paramagnetic LuFe$_2$Zn$_{20}$, to compare with the published data on YFe$_2$Zn$_{20}$ and evaluate the steric effects on the hyperfine parameters, and on the heavy fermion member of the family: YbFe$_2$Zn$_{20}$.

\section{Experimental methods}

Single crystalline samples of  {\it R}Fe$_2$Zn$_{20}$ ({\it R} = Lu, Yb, Gd) were grown out of excess Zn using standard solution growth techniques.\cite{can92a} Initial ratios of starting elements (R:Fe:Zn) were 2:4:94.   The constituent elements were placed in an alumina crucible, sealed in a quartz ampule under $\sim 1/3$ atmosphere Ar, heated in a box furnace to 900-1150$^{\circ}$ C and then slowly cooled to 600$^{\circ}$ C over  $\sim 100$ h. More details on single crystal growth of these materials are provided in Refs. \cite{jia07a,tor07a}.

M\"ossbauer spectroscopy measurements were performed using a SEE Co. conventional constant acceleration type spectrometer in transmission geometry with an $^{57}$Co(Rh) source, which had an initial (8-9 months before the measurements) intensity 50 mCi, kept at room temperature. For the absorber, the {\it R}Fe$_2$Zn$_{20}$ ({\it R} = Lu, Yb, Gd) single crystals were powdered using mortar and pestle and mixed with a ZG grade BN powder to ensure homogeneity.  The absorber holder comprised two nested white Delrin cups.  The absorber holder was locked in a thermal contact with a copper block with a temperature sensor and a heater, and aligned with the $\gamma$ - source and detector.   The absorber was cooled to a desired temperature using a Janis model SHI-850-5 closed cycle refrigerator (with vibration damping). The driver velocity was calibrated by $\alpha$-Fe foil and all isomer shifts (IS) are quoted relative to the $\alpha$-Fe foil at room temperature. A commercial software package MossWinn \cite{MW} were used to analyze the M\"ossbauer spectra in this work.

\section{Results and discussion}

\subsection{LuFe$_2$Zn$_{20}$ and YbFe$_2$Zn$_{20}$ }

\begin{figure}[htbp]
\begin{center}
\includegraphics[angle=0,width=120mm]{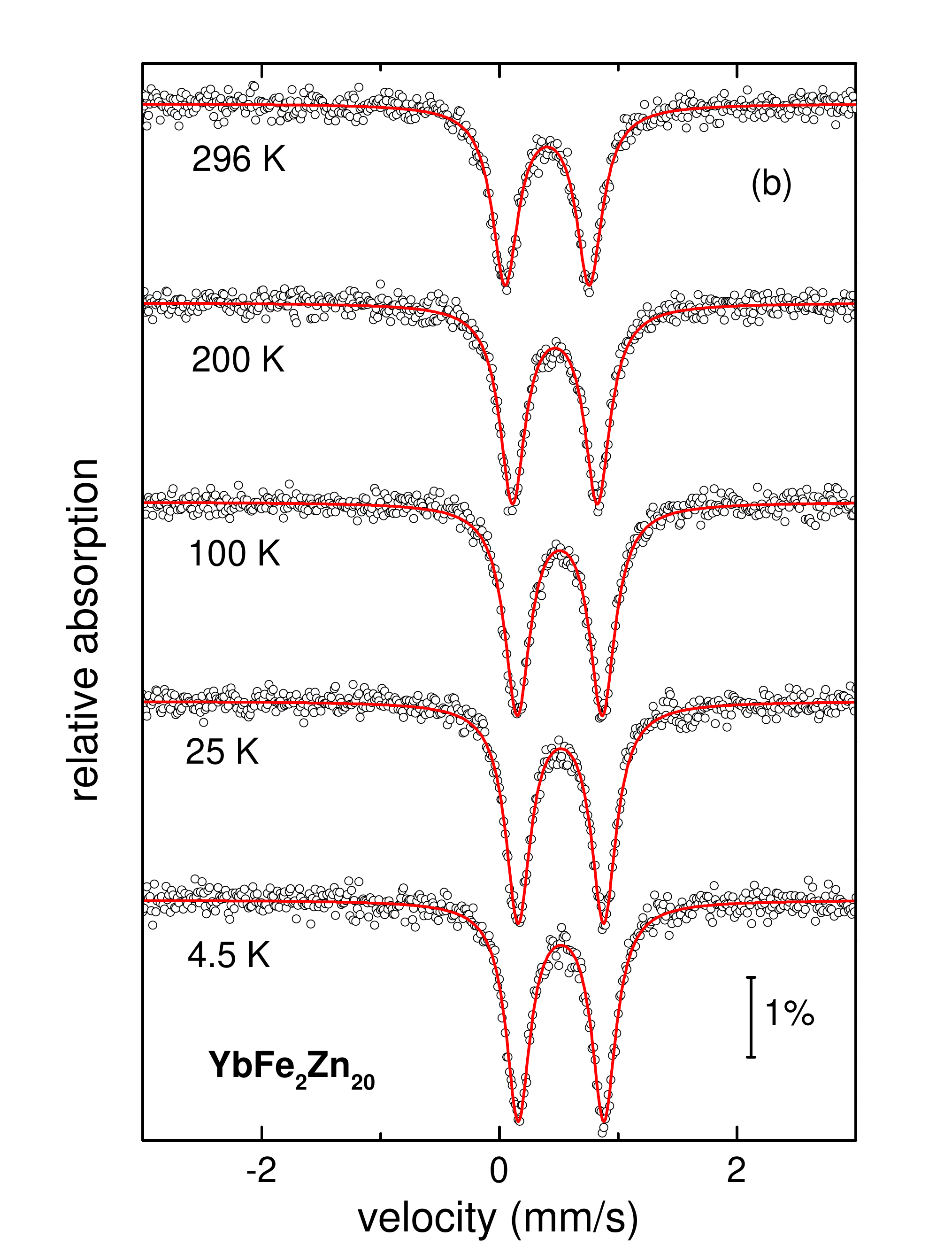}
\end{center}
\caption{(Color online) $^{57}$Fe M\"ossbauer spectra of (a) LuFe$_2$Zn$_{20}$ and (b) YbFe$_2$Zn$_{20}$ at selected temperatures. Circles - experimental data, lines - fits.} \label{F1}
\end{figure}

The Fe atoms occupy a single $16d$ site with $\bar{3}m$ (trigonal) point symmetry in the cubic CeCr$_2$Al$_{20}$ type structure ($Fd\bar{3}m$, $Z = 8$). \cite{nas97a} The non-spherical environment of the Fe atoms (Fig. \ref{Fe}) ensures a nonzero electric field gradient tensor (EFG) at the $^{57}$Fe site. Indeed, the $^{57}$Fe M\"ossbauer spectra of LuFe$_2$Zn$_{20}$ and YbFe$_2$Zn$_{20}$ taken at different temperatures (Fig. \ref{F1}) show similar spectral shapes with a clear quadrupole splitting. Each of these spectra can be fitted with a single doublet, consistent with Fe occupying a single, unique crystallographic site. The spectra are very similar for both compounds and are very weakly temperature-dependent.

\begin{figure}[htbp]
\begin{center}
\includegraphics[angle=0,width=120mm]{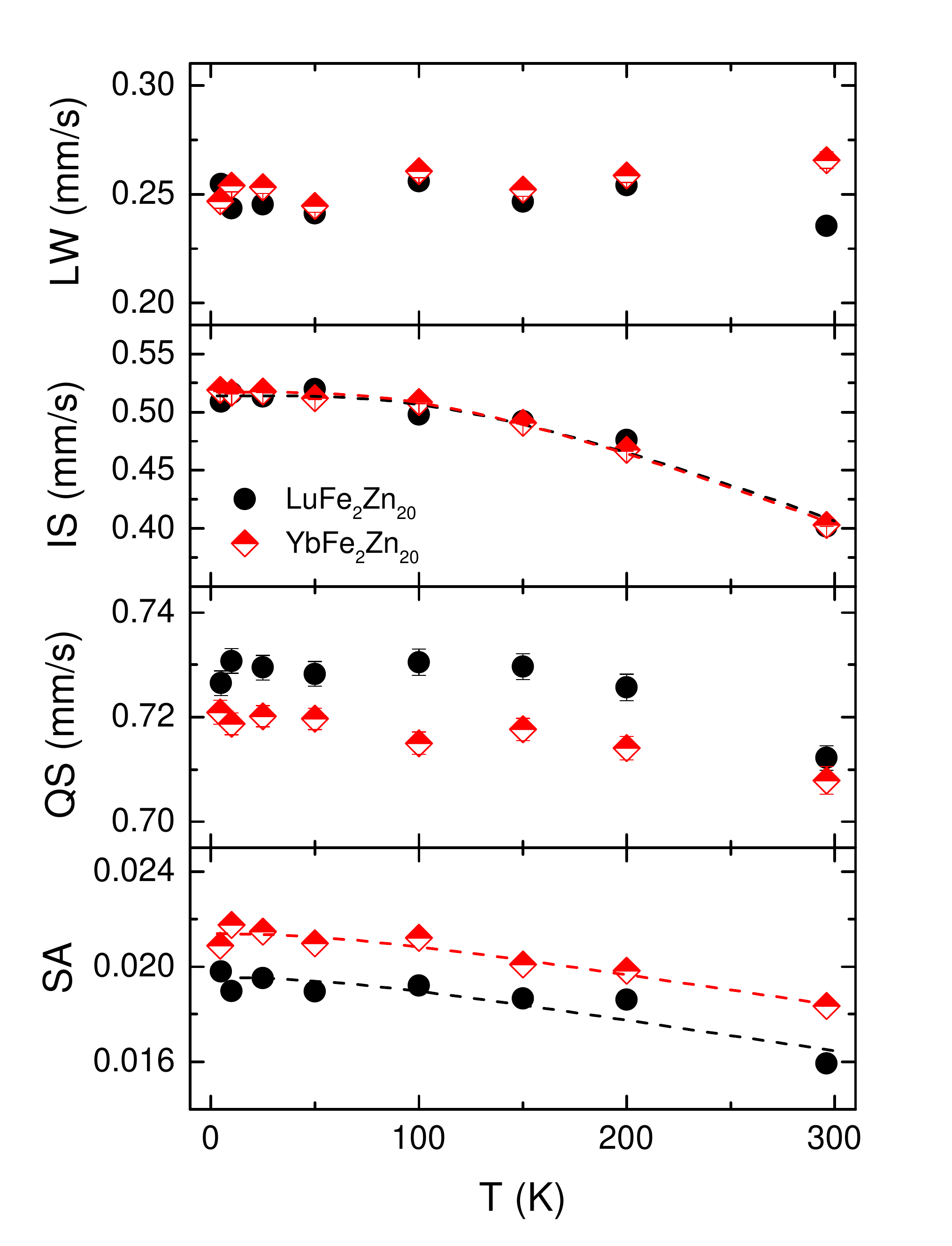}
\end{center}
\caption{(Color online) Temperature dependence of the hyperfine parameters obtained from $^{57}$Fe M\"ossbauer spectra of  LuFe$_2$Zn$_{20}$ (filled circles) and YbFe$_2$Zn$_{20}$ (half filled diamonds). LW - linewidth, IS - isomer shift, QS - quadrupole splitting, SA - normalized spectral area. Dashed lines - Debye fits as described in the text.} \label{F2}
\end{figure}

The temperature dependent hyperfine parameters for LuFe$_2$Zn$_{20}$ and YbFe$_2$Zn$_{20}$  are shown in Fig. \ref{F2}. These parameters are very similar for both compounds. The linewidth is basically temperature independent. The isomer shift increases with decrease in temperature. The main contribution to the temperature dependence of the isomer shift comes from the second order Doppler shift and is usually described by the Debye model: 
\begin{equation}
IS(T)=IS(0)-\frac{9}{2}\frac{k_BT}{Mc}\left (\frac{T}{\Theta_D}\right )^3\int_0^{\Theta_D/T}\frac{x^3dx}{e^x-1},
\end{equation}
where $c$ is the velocity of light,  $M$ is the mass of the $^{57}$Fe nucleus, and $IS(0)$ is is the temperature-independent part, i.e. the chemical shift. The obtained values of the Debye temperature, $\Theta_D$ are $524 \pm 47$ K and $487 \pm 17$ K for LuFe$_2$Zn$_{20}$ and YbFe$_2$Zn$_{20}$ respectively. 

The quadrupole splitting increases slightly, by $\sim 1.5 - 3 \%$ on cooling from room temperature to the base temperature. 

Temperature dependence of the relative spectral area (SA) in absence of phase transitions can be described within the Debye model as well.  

\begin{equation}
  f=exp\left [\frac {-3E_{\gamma}^2}{k_B\Theta_DMc^2} \left \{ \frac{1}{4}+\left (\frac{T}{\Theta_D}\right )^2 \int_0^{\Theta_D/T}\frac{xdx}{e^x-1} \right \} \right ],
\end{equation}
where $f$ is the recoilless fraction, which is proportional to the spectral area for thin sample and E$_{\gamma}$ is the $\gamma$-ray energy. 
The values of $\Theta_D$ obtained from the fits are $410 \pm 32$ K and $433 \pm 25$ K for LuFe$_2$Zn$_{20}$ and YbFe$_2$Zn$_{20}$ respectively.

\subsection{GdFe$_2$Zn$_{20}$}

\begin{figure}[htbp]
\begin{center}
\includegraphics[angle=0,width=120mm]{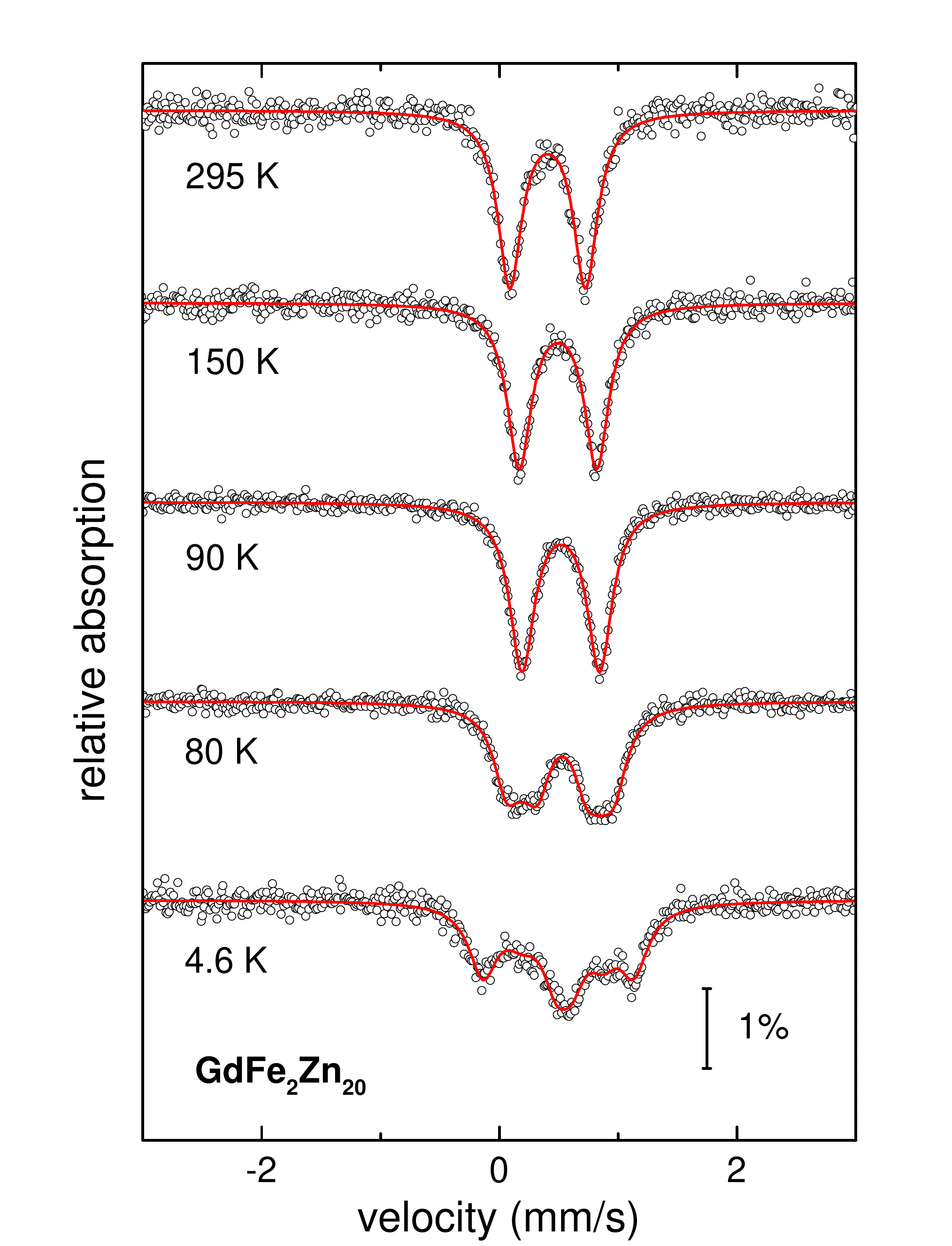}
\end{center}
\caption{(Color online) $^{57}$Fe M\"ossbauer spectra of GdFe$_2$Zn$_{20}$ at selected temperatures. Circles - experimental data, lines - fits.} \label{spectraGd}
\end{figure}

 The $^{57}$Fe M\"ossbauer spectra of GdFe$_2$Zn$_{20}$, taken at different temperatures, are presented in Fig. \ref{spectraGd}. Whereas at high temperatures the observed spectra are doublets and are very similar to those measured for  LuFe$_2$Zn$_{20}$ and YbFe$_2$Zn$_{20}$ (Fig. \ref{F1}), at the base temperature the spectrum splits in six lines indicating the presence of magnetic hyperfine field at the Fe site, in agreement with ferromagnetic ground state of GdFe$_2$Zn$_{20}$ proposed based on the bulk magnetic measurements. \cite{jia07a}  The low temperature spectra for GdFe$_2$Zn$_{20}$ depend on the electric field gradient and magnetic hyperfine field, $B_{hf}$, at the $^{57}$Fe nuclei, as well as on the angle between the principal axis of the EFG tensor and the direction of the  $B_{hf}$. The "mixed Q + M static Hamiltonian (powder)" model (the $z$-axis is parallel to the principal axis of the EFG temsor, the static Hamiltonian includes magnetic and quadrupole interactions with an arbitrary relative orientation) of the MossWinn \cite{MW} program was used to analyze these spectra. Due to the relative symmetry of the environment of the Fe atom in the lattice (see Fig. \ref{Fe}), the EFG assymetry parameter, $\eta = (V_{xx} - V_{yy})/V_{zz}$ and the azimuthal angle of the hyperfine field, $\alpha$ were fixed at zero for the fits.

\begin{figure}[htbp]
\begin{center}
\includegraphics[angle=0,width=120mm]{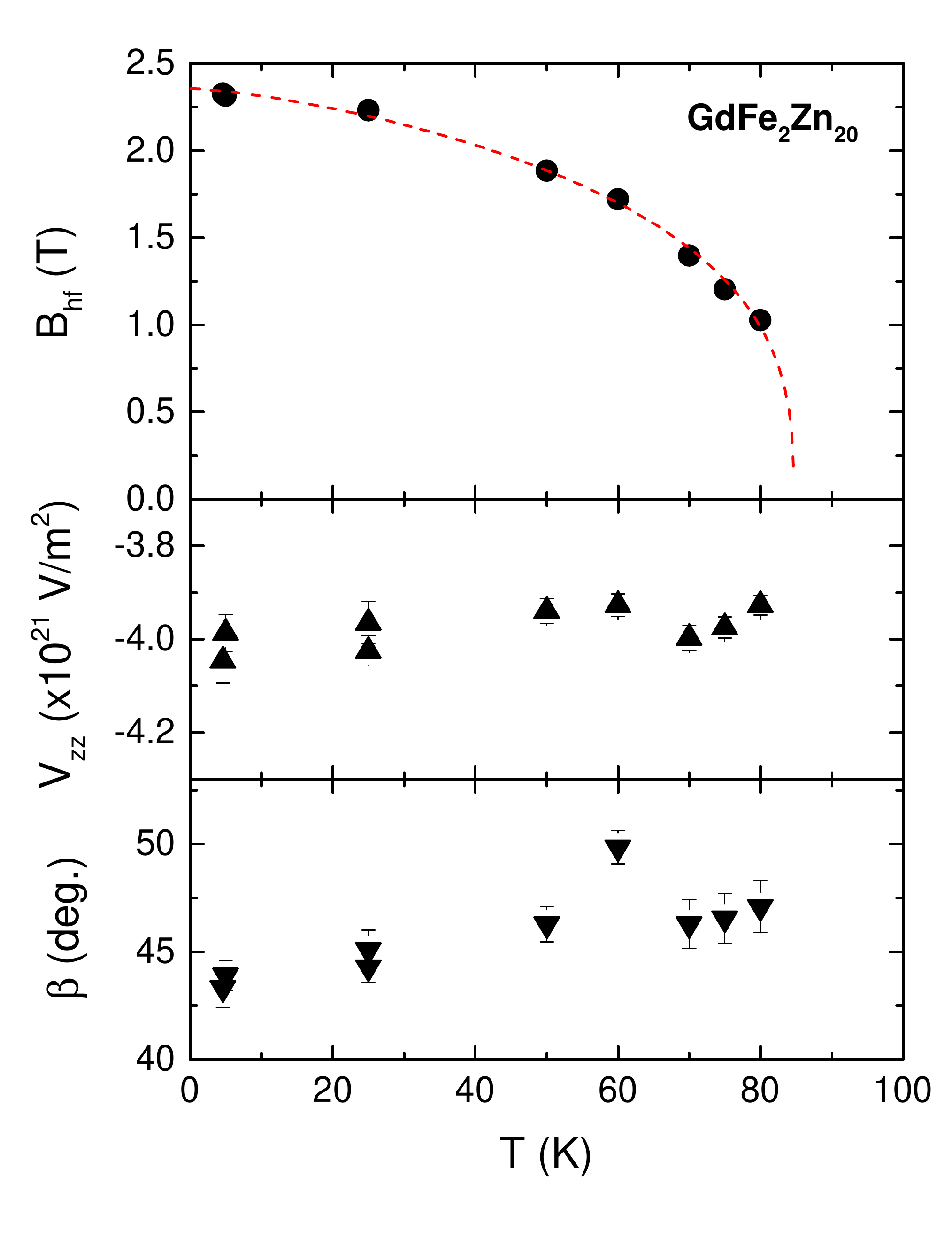}
\end{center}
\caption{(Color online) Top panel: magnetic hyperfine field, $B_{hf}(T)$,  for GdFe$_2$Zn$_{20}$. Symbols - data,  dashed line - fit to Curie - Bloch equation - see the text. Middle panel: $V_{zz}$  componen of the telectric field gradient tensor as a function of temperature. Bottom panel: temperature dependence of the angle between $V_{zz}$ and the direction of the $B_{hf}$.} \label{Bhf}
\end{figure}

Hyperfine field at the Fe site, $B_{hf}$ as a function of temperature is shown in Fig. \ref{Bhf}. At the base temperature the value of $B_{hf}$ is close to 2.4 T. It has been observed \cite{pre62a} that temperature dependence of the hyperfine field follows closely (even if not exactly) the saturation magnetization curve. We can fit  $B_{hf}(T)$ data with the function based on the Curie-Bloch equation \cite{eva15a} for saturation magnetization.

\begin{equation}
 B_{hf} = B_{hf}(0)\left [1-\left (\frac{T}{T_c}\right ) ^{\alpha}\right ]^{\beta},
\end{equation}
where $\alpha$ is an empirical constant and $\beta$ is a critical exponent. The value $\beta = 1/3$ \cite{eva15a} was used in the fit. As the result of the fit the following values were obtained: $B_{hf}(0) = 2.36 \pm 0.03$ T, $T_C = 84.6 \pm 0.8$ K, $\alpha = 1.4 \pm 0.1$. The bulk measurements of GdFe$_2$Zn$_{20}$ \cite{jia07a,jia08a} yield  a $T_C = 86 \pm 2$ K value of the Curie temperature, that is consistent with the results of the fit and with our observation of a magnetic sextet in the 80 K spectrum and a non-magnetic doublet in the 90 K spectrum (Fig. \ref{spectraGd}). 

The values of the $V_{zz}$ component of the EFG tensor and of the polar angle $\beta$ between the directions of $V_{zz}$ and $B_{hf}$ (Fig. \ref{Bhf}) show only insignificant change between $\sim 4.5$ K and 80 K.

\begin{figure}[htbp]
\begin{center}
\includegraphics[angle=0,width=120mm]{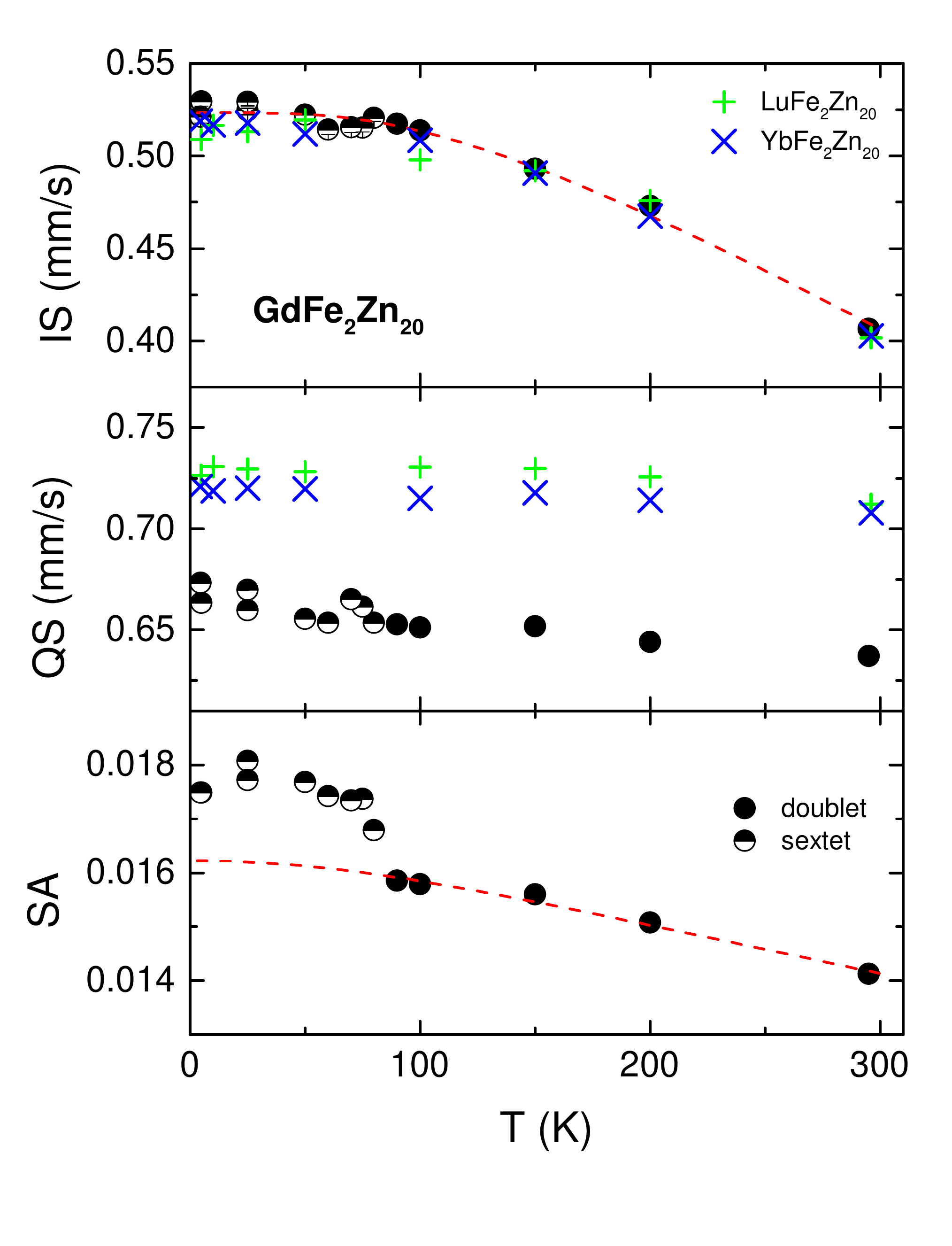}
\end{center}
\caption{(Color online) Temperature dependence of the hyperfine parameters obtained from $^{57}$Fe M\"ossbauer spectra of  GdFe$_2$Zn$_{20}$.   IS - isomer shift, QS - quadrupole splitting, SA - normalized spectral area. Dashed lines - Debye fits as described in the text. For $IS(T)$ and $QS(T)$ data for LuFe$_2$Zn$_{20}$ ($+$) and YbFe$_2$Zn$_{20}$ ($\times$) are plotted for comparison.} \label{fitGd}
\end{figure}

The temperature dependent hyperfine parameters of GdFe$_2$Zn$_{20}$ are shown in Fig. \ref{fitGd}. The values of $IS(T)$, $QS(T)$ and the general behavior of $SA(T)$ are consistent with those found for LuFe$_2$Zn$_{20}$ and YbFe$_2$Zn$_{20}$. The overall behavior of the isomer shift  can be described by the Debye model with $\Theta_D = 465 \pm 23$ K, similarly, above the magnetic transition, the Debye model with $\Theta_D = 450 \pm 14$ K can be used to describe the temperature dependent relative spectral area. It should be noted that an increase in the spectral area (beyond to what is expected in the Debye model) is apparent on decrease of temperature below the transition. Since significant change in phonon spectra is not anticipated in the ferromagnetic state, and since the $IS(T)$ behavior in GdFe$_2$Zn$_{20}$ is very similar to those in  LuFe$_2$Zn$_{20}$ and YbFe$_2$Zn$_{20}$, probably some unknown artifacts contribute to the curious  $SA(T)$ behavior shown in Fig. \ref{fitGd}.  The quadrupole splitting is only weakly temperature dependent both above and below the magnetic transition. $QS(295$K$) = 0.637 \pm 0.005$ mm/s that is $\sim 10 \%$ lower than the room temperature values for  LuFe$_2$Zn$_{20}$ and YbFe$_2$Zn$_{20}$ ($0.712 \pm 0.002$ and $0.708 \pm 0.003$, respectively).

\subsection{Discussion}

The values and the temperature dependence of the isomer shift for all three compounds studied in this work are very similar and are consistent with the results reported for YFe$_2$Zn$_{20}$ and DyFe$_2$Zn$_{20}$. \cite{tam13a} The Debye temperature, evaluated from the $IS(T)$ data is $\sim 450 - 500$ K for all of them. Similar but slightly lower values of Debye temperature, $\sim 410 - 450$ K, were estimated for {\it R}Fe$_2$Zn$_{20}$ ({\it R} = Lu, Yb, Gd) from the temperature dependence of the relative spectral area. This discrepancy may be explained by the fact the area reflects the average mean-square displacements, whereas $IS$ is related to the mean-square velocity of the Mössbauer atom. Both quantities may respond in different ways to the lattice anharmonicities. It has to be mentioned that the $\Theta_D$ values inferred from the temperature dependence of the hyperfine parameters above are significantly higher than $\Theta_D \approx 340$ K evaluated for  YFe$_2$Zn$_{20}$ and  LuFe$_2$Zn$_{20}$ from the low temperature specific heat capacity. \cite{jia09a} This discrepancy might be related to the fact that the $\Theta_D$ estimates from the temperature dependent M\"ossbauer spectra reflect, to a large extent, the phonon frequencies related to the M\"ossbauer atom ($^{57}$Fe), whereas specific heat capacity involves more uniform average. Additionally, other, e.g. electronic, contributions might play a role in the temperature evolution of the $IS$ and $SA$ in this family of materials with nearly ferromagnetic Fermi liquid matrix.

The $^{57}$Fe hyperfine parameters of the Yb-based heavy fermion YbFe$_2$Zn$_{20}$ and nearly ferromagnetic Fermi liquid compound LuFe$_2$Zn$_{20}$ are virtually indistinguishable. Possibly the effects of additional electron correlations below the Kondo temperature of $\sim 33$ K in YbFe$_2$Zn$_{20}$, \cite{tor07a} as probed on the Fe site, are small in comparison with the  effects of strongly correlated Fermi liquid governed by the transition metal density of states.\cite{jia07a}

\begin{figure}[htbp]
\begin{center}
\includegraphics[angle=0,width=120mm]{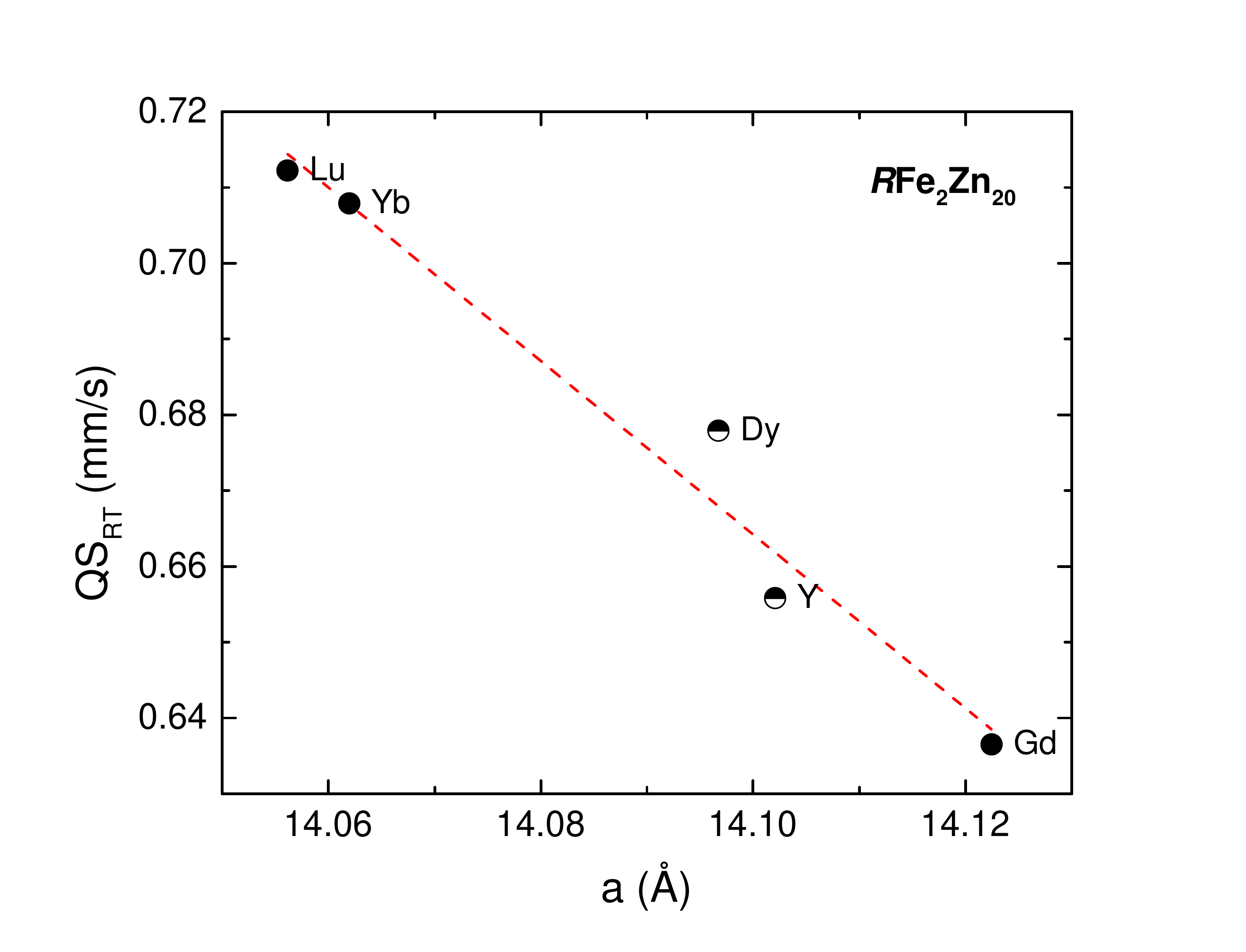}
\end{center}
\caption{(Color online) Quadrupole splitting at room temperature as a function of the $a$ lattice parameter. Filled circles - this work, half-filled circles - from Ref. \cite{tam13a}. Dashed is a guide to the eye. Values of the lattice parameter $a$ were taken from Ref. \cite{jia09a}} \label{QSRT}
\end{figure}

The quadrupole splitting values are governed by the local environment of the Fe atoms (Fig. \ref{Fe}). To follow the geometrical changes in this environment with the change of rare earth ionic radius or the lattice parameters is a cumbersome task and does require a detailed knowledge of atomic coordinates for each compound. Experimentally though, it appears that the $QS$ at room temperature decreases as a function of the $a$-lattice parameter in close to linear fashion (Fig. \ref{QSRT}) suggesting the trend of the Fe environment to become more isotropic with increase of $a$. This observation is consistent with a very simple notion the average electric field on the Fe site becoming "more spherical" with point charges of the local environment moving further apart. 

The ferromagnetic transition in GdFe$_2$Zn$_{20}$ is seen in $^{57}$Fe M\"ossbauer spectra as appearance of a sextet (Fig. \ref{spectraGd}), additionally, there is an increase in the relative spectral area near Curie temperature (Fig. \ref{fitGd}). The value of the hyperfine field for $T \to 0$ is  $B_{hf}(0) = 2.36 \pm 0.03$ T, that is $\sim 25 \%$  higher than $\sim 1.84$ T measured in DyFe$_2$Zn$_{20}$. \cite{tam13a} It is possible that the crystal electric field effect present in DyFe$_2$Zn$_{20}$ and absent in GdFe$_2$Zn$_{20}$ should be considered in the analysis of this difference.

The temperature dependence of the  $^{57}$Fe site hyperfine field in GdFe$_2$Zn$_{20}$  allows us to address the question of interpretation of the strong deviation of the temperature dependence of magnetic susceptibility from the Curie-Weiss law  between $T_C$ and  $\sim 250$ K, discussed in Ref. \cite{jia08a}. Absence of hyperfine field on the  $^{57}$Fe site at and above 90 K, argues against the picture of formation of magnetic droplets consisting of the Gd$^{+3}$ local moments and oppositely polarized electron cloud, leaving the hypothesis of the temperature dependent coupling between Gd$^{3+}$ local moments as more viable.

\section{Summary}

To summarize, we performed measurements of $^{57}$Fe M\"ossbauer spectra for {\it R}Fe$_2$Zn$_{20}$ ({\it R} = Lu, Yb, Gd) from $\sim 4.5$ K to room temperature. The isomer shift values are very similar for all three compounds, and give an estimate of $\Theta_D$ in the range of 480-500 K. The values of quadrupole splitting at room temperature change with the lattice constant in a linear fashion. Ferromagnetic order is seen as a sextet in the spectra. The  $B_{hf}(T)$ data can be fitted with the function based on the Curie-Bloch equation for saturation magnetization that yields the value of $B_{hf}$ at $T \to 0$ of $\sim 2.4$ T. No features in the M\"ossbauer spectra were associated with the Yb-based heavy fermion behavior in the YbFe$_2$Zn$_{20}$ at low temperatures.

This study suggests that further work on understanding of the observed trend in $QS$ within this family as well as on elucidation of the relation of the measured hyperfine field and bulk magnetic properties of the compounds in the {\it R}Fe$_2$Zn$_{20}$ family is desired.

\ack

\begin{figure}[htbp]
\begin{center}
\includegraphics[angle=0,width=150mm]{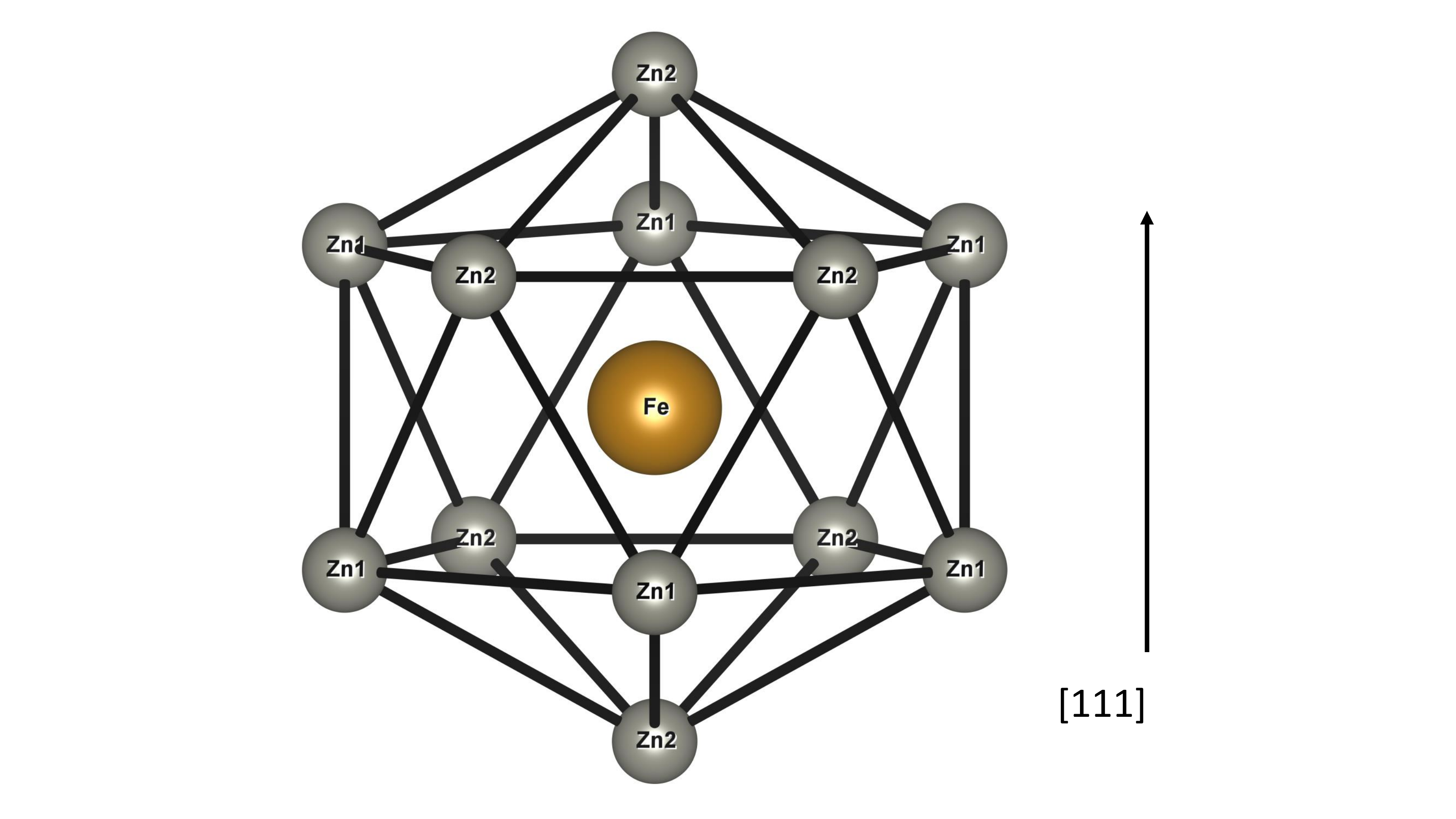}
\end{center}
\caption{(Color online) Local environment of the Fe atoms in  {\it R}Fe$_2$Zn$_{20}$. (After Ref. \cite{nas97a}). } \label{Fe}
\end{figure}

We are very grateful to Dominic Ryan, for the critical reading of the manuscript and many useful suggestions. We thank Shuang Jua for synthesis of some early samples of {\it R}Fe$_2$Zn$_{20}$ and  Udhara Kaluarachchi for help in preparation of the Fig. \ref{Fe}. Work at the Ames Laboratory was supported by the US Department of Energy, Basic Energy Sciences, Division of Materials Sciences and Engineering under Contract No. DE-AC02-07CH11358. X.M. was supported in part by the China Scholarship Council.

\clearpage

\clearpage

\begin{figure}[htbp]
\begin{center}
\includegraphics[angle=0,width=120mm]{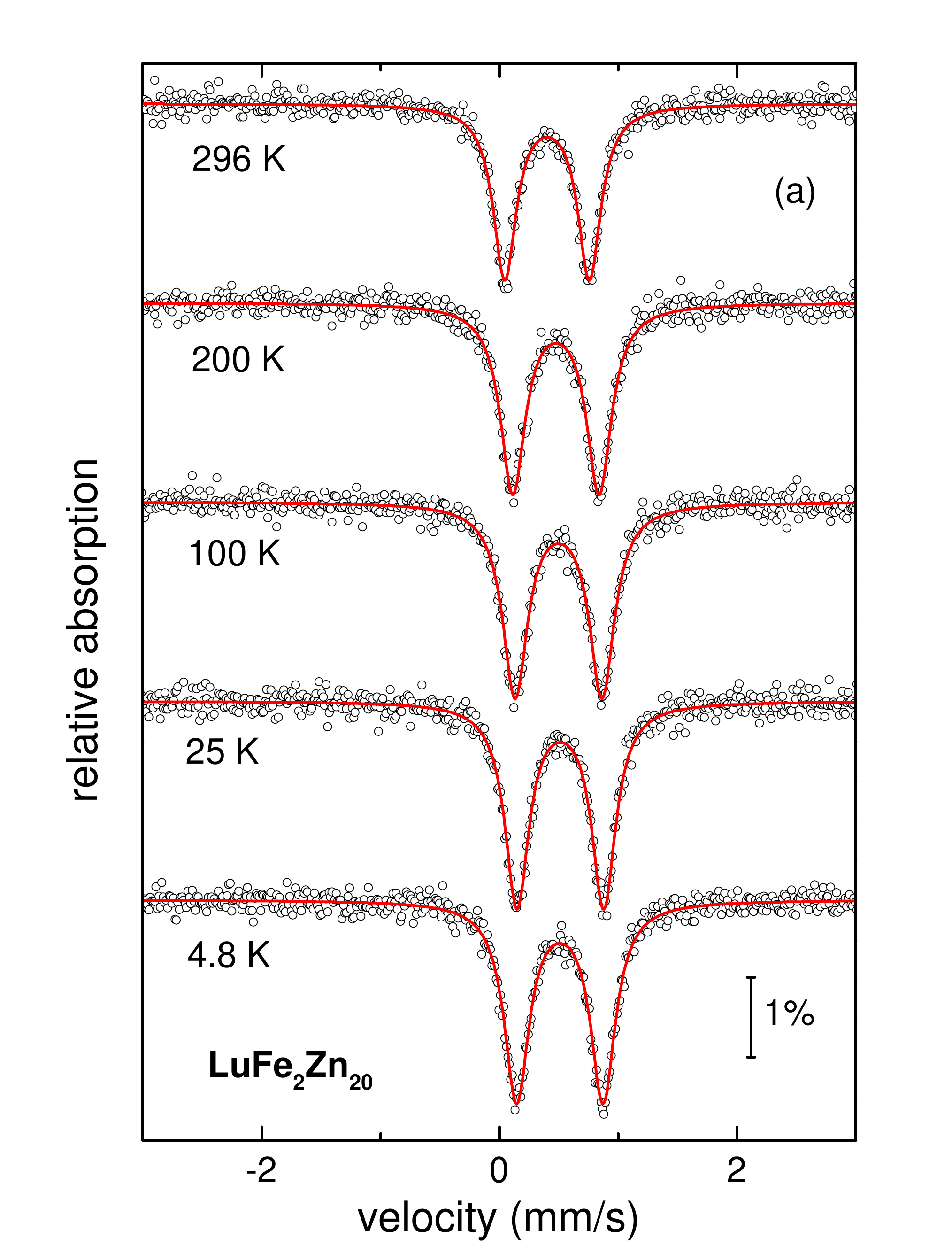}
\end{center}
\end{figure}

\clearpage

\clearpage

\clearpage

\clearpage

\clearpage

\clearpage

\section*{References}

\begin{thebibliography}{99}

\bibitem{nas97a}
 Nasch T, Jeitschko W, and Rodewald U C 1997 \textit{Z. Naturforsch., B: Chem. Sci.} \textbf{52} 1023.

\bibitem{jia07a}
Jia S, Bud'ko S L, Samolyuk G D, and Canfield P C 2007 \textit{Nat. Phys.} \textbf{3} 334.

\bibitem{tor07a}
Torikachvili M S, Jia S, Mun E D, Hannahs S T, Black R C, Neils W K, Martien D, Bud'ko S L, and Canfield P C 2007  \textit{Proc. Natl. Acad. Sci. U.S.A.} \textbf{104} 9960.

\bibitem{jia09a}
Jia Shuang, Ni Ni,  Bud'ko S L, and Canfield P C 2009 \textit{Phys. Rev. B} \textbf{80} 104403.

\bibitem{jia07b}
Jia Shuang, Ni Ni,  Bud'ko S L, and Canfield P C 2007 \textit{Phys. Rev. B} \textbf{76} 184410.

\bibitem{jia08a}
Jia Shuang, Ni Ni, Samolyuk G D, Safa-Sefat A, Dennis K, Ko Hyunjin,  Miller G J, Bud'ko S L, and Canfield P C 2008 \textit{Phys. Rev. B} \textbf{77} 104408.

\bibitem{tia10a}
Tian W, Christianson A D, Zarestky J L, Jia S, Bud'ko S L, Canfield P C, Piccoli P M B, and  Schultz A J 2010 \textit{Phys. Rev. B} \textbf{81} 144409.

\bibitem{isi13a}
Isikawa Yosikazu, Mizushima Toshio, Miyamoto Souta, Kumagai Keigou, Nakahara Mako, Okuyama Hiroaki, Tayama Takashi, Kuwai Tomohiko, and Lejay Pascal 2013 \textit{J. Korean Phys. Soc.} \textbf{63} 644.

\bibitem{yaz14a}
Yazici D,  White B D, Ho P-C, Kanchanavatee N, Huang K, Friedman A J, Wong A S, Burnett V W, Dilley N R, and Maple M B 2014 \textit{Phys. Rev. B} \textbf{90} 144406.

\bibitem{wan10a}
Wang C H, Christianson A D, Lawrence J M, Bauer E D, Goremychkin E A,  Kolesnikov A I, Trouw F, Ronning F,  Thompson J D,  Lumsden M D, Ni N, Mun E D, Jia S,  Canfield P C, Qiu Y, and Copley J R D 2010 \textit{Phys. Rev. B} \textbf{82} 184407.

\bibitem{tam13a}
Tamura Ichiro, Isikawa Yosikazu, Mizushima Toshio, and Miyamoto Souta 2013 \textit{J. Phys. Soc. Jpn.}  \textbf{82} 114703.

\bibitem{gut11a}
G\"utlich P, Bill E, and Trautwein A X 2011 \textit{M\"ossbauer Spectroscopy and Transition Metal Chemistry} (Berlin, Heidelberg: Springer-Verlag)

\bibitem{can92a}
 Canfield P C, Fisk Z 1992 \textit{Phil Mag B} \textbf{65} 1117.

\bibitem{MW}
Klencs\'ar Z, \textit{MossWinn 4.0 Pre}.

\bibitem{pre62a}
Preston R S, Hanna S S, and Heberle J 1962 \textit{Phys. Rev.} \textbf{128} 2207.

\bibitem{eva15a}
Evans R F L, Atxitia U, and Chantrell R W 2015 \textit{Phys. Rev. B} \textbf{91} 144425.




\end{thebibliography}
\end{document}